\documentclass[aps,prd,preprintnumbers,superscriptaddress,nofootinbib,amsmath,amssymb,showpacs,10pt]{revtex4}
\usepackage{graphicx}
\usepackage{dcolumn}
\usepackage{bm}
\usepackage{amsmath}
\usepackage{color}
\input{colordvi.tex}

\newcommand{\gsim}{~\mbox{\raisebox{-1.0ex}{$\stackrel{\textstyle >}
{\textstyle \sim}$ }}}
\newcommand{\lsim}{~\mbox{\raisebox{-1.0ex}{$\stackrel{\textstyle <}
{\textstyle \sim}$ }}}

\newcommand{\beq}{\begin{equation}}
\newcommand{\eeq}{\end{equation}}
\newcommand{\beqa}{\begin{eqnarray}}
\newcommand{\eeqa}{\end{eqnarray}}

\begin{document}

\begin{flushleft}
RESCEU-52/14 \\
YITP-14-100
\end{flushleft}

\title{Effects of thermal fluctuations on thermal inflation}

\author{Takashi Hiramatsu}
\email[Email: ]{hiramatz"at"yukawa.kyoto-u.ac.jp}
\affiliation{Yukawa Institute for Theoretical Physics, Kyoto University, Kyoto 606-8502, Japan}

\author{Yuhei Miyamoto}
\email[Email: ]{miyamoto"at"resceu.s.u-tokyo.ac.jp}
\affiliation{ Department of Physics, Graduate School of Science, \\
The University of Tokyo, Tokyo 113-0033, Japan}
\affiliation{Research Center for the Early Universe (RESCEU), \\
Graduate School of Science, The University of Tokyo, Tokyo 113-0033, Japan}

\author{Jun'ichi Yokoyama}
\email[Email: ]{yokoyama"at"resceu.s.u-tokyo.ac.jp}
\affiliation{Research Center for the Early Universe (RESCEU), \\
Graduate School of Science, The University of Tokyo, Tokyo 113-0033, Japan}
\affiliation{Kavli Institute for the Physics and Mathematics of the Universe (Kavli IPMU),
WPI, TODIAS, The University of Tokyo, Kashiwa, Chiba, 277-8568, Japan}


\begin{abstract}
The mechanism of thermal inflation, a relatively short period of
accelerated expansion after primordial inflation, is a desirable
ingredient for a certain class of particle physics models if they are
not to be in contention with the cosmology of the early Universe.
Though thermal inflation is most simply described in terms of a thermal
effective potential, a thermal environment also gives rise to thermal
fluctuations that must be taken into account.  We numerically study the
effects of these thermal fluctuations using lattice simulations.  We
conclude that though they do not ruin the thermal inflation scenario,
the phase transition at the end of thermal inflation proceeds through phase mixing
 and is therefore not accompanied by the formations of
bubbles nor appreciable amplitude of gravitational waves.

\end{abstract}
\maketitle

\section{Introduction}
The idea of the inflationary Universe \cite{inflation} is now a key part
of the standard model of cosmology.  The primordial period of
accelerated expansion at the beginning of the Universe provides not only
a solution to the flatness and horizon problems, but also the initial
density fluctuations that seed the formation of large-scale structure.

It has been claimed that a period of accelerated expansion has the
potential to reconcile a certain class of particle physics models with
cosmology.  The gravitino, a fermionic partner of the graviton with spin
$3/2$, appears in the theory of supergravity.  Its number density per
comoving volume is proportional to the reheating temperature after
inflation \cite{gravitinoproblem}.  Therefore, if the reheating
temperature is high, the gravitinos are abundantly produced.  
The lifetime of the gravitino is estimated as
$\tau \sim 8\pi M_{\rm Pl}^2/m_{3/2}^3\sim 10^{5}\,{\rm sec}$
if the gravitino mass takes a value $m_{3/2}=10^3 {\rm GeV}$
with $M_{\rm Pl}=2.4\times 10^{18}\, {\rm GeV}$ being the reduced Planck mass.
Namely,
they decay after Big-Bang Nucleosythesis (BBN) due to their very weak
interactions.  Subsequently, the decay products of gravitinos spoil the
light elements after BBN.  This is called the gravitino problem.
The scalar fields called moduli, with Planck-suppressed couplings, are
also dangerous in a similar way \cite{moduliproblem}.  They start to
oscillate when the Hubble parameter becomes as small as their mass and
soon dominate the Universe, since the initial amplitude of such
oscillations is expected to be on the order of $M_{\rm Pl}$.  Driven by
the coherent oscillations of the moduli fields the Universe evolves like
a matter-dominated one, until the moduli decay to reheat the Universe.
The moduli fields are coupled very weakly with other fields, and as
a result of their long lifetime the reheating temperature is
so low that BBN does not work.
Furthermore, in Ref. \cite{Kawasaki:1997ah} 
it is shown that the energy density of moduli
is also constrained by X($\gamma$)-ray observations,
requiring that the theoretical prediction does not exceed the observed backgrounds.
One can dilute the moduli fields by assuming a short, low-energy
inflationary period after the moduli begin oscillating at $H\approx
m_{\rm moduli}$ \cite{Yamamoto:1985rd,Lyth:1995hj, moduliproblem_TI}.  
This type of temporally short inflationary period is
called thermal inflation, and is driven by a scalar field with almost
flat potential called the flaton \cite{Yamamoto:1985rd,Lyth:1995hj}.  
In a similar way,
thermal inflation can also evade the gravitino problem
\cite{gravitinoproblem} by diluting them after their generation.  In
summary, thermal inflation is needed to dilute the unwanted relics
formed after primordial inflation in a similar way that primordial
inflation can solve the monopole problem of the big bang model.

Thermal inflation has been studied by many authors 
due to its other interesting properties.  First, it is related to gravitational
waves.  A period of accelerated expansion after the generation of tensor
perturbations in the primordial inflationary period leads to their
dilution \cite{diluteGWs}.  This is a non-negligible effect that must be
taken into account when determining the value of the primordial
tensor-to-scalar ratio and constraining models of inflation using
observations.  In addition, the collision of bubbles created at the end
of thermal inflation can give rise to gravitational waves
\cite{Easther:2008sx,generateGWs}.
Second, thermal inflation provides a mechanism for baryogenesis.  Though
it washes out the baryon number generated before thermal inflation, we
can consider mechanisms for generating baryon asymmetry at the end of
thermal inflation \cite{baryogenesis}.
Third,
effects of thermal inflation on the primordial density fluctuations are studied in Ref. \cite{Kawasaki:2009hp}.

In a similar way as with primordial inflation, the mechanism of thermal
inflation is often described in terms of an effective potential.  A key
difference with most models of primordial inflation, however, is that
there exists a radiation bath during thermal inflation.  Interactions
with particles in the thermal bath lead to thermal corrections to the
flaton potential, which creates a small dip at the origin of the flaton
potential.  Thermal inflation is driven by the potential energy of
the flaton at the origin and we usually assume it ends through a
first-order phase transition.

Though the existence of a thermal bath is necessary for thermal
inflation to occur, it also leads to thermal fluctuations that affect
the dynamics of the flaton field.  Since these effects are not accounted
for in the effective potential approach, we incorporate the effect of
thermal fluctuations separately.  In this paper we consider two phases
which are relevant to the thermal inflation scenario.  The first phase
is before the beginning of thermal inflation.  If in some spatial
regions the flaton value is kept large even when the Universe cools,
thermal inflation never begins.
The second phase is the end of thermal inflation.  If thermal inflation
ends with a first-order phase transition, bubbles are generated and
their collisions induce gravitational waves.  Therefore, in order to
predict gravitational-wave observables, it is important to study how
thermal inflation ends with thermal fluctuations taken into account.

This paper is organized as follows.  In Section \ref{TIS}, we take a
brief look at the thermal inflation scenario.  Though it is often
described in terms of an effective potential, we consider the flaton
dynamics based on the effective action in Section \ref{fdtb}.  We study
the flaton dynamics further in detail by performing lattice simulations,
whose setup is summarized in Section \ref{simulationsetup}, and discuss
the results in Section \ref{RNS}.  In Section \ref{conclusion}, we
summarize the implications of our study for the thermal inflation
scenario.

\section{Scenario of Thermal Inflation \label{TIS}}
We briefly review the scenario of thermal inflation in this section.  In
considering the dynamics of thermal inflation, we often use the thermal
effective potential.  Since thermal inflation occurs
after primordial inflation and reheating, there is a hot thermal bath
and interactions between the flaton and the fields in the bath lead to
thermal corrections to the flaton potential.  The flaton is kept at the
origin of the potential owing to this correction and the potential
energy at the origin drives thermal inflation.  One example of the
flaton potential at zero temperature is
\begin{equation}
V_0(\phi)=V_{\rm TI}-\frac{1}{2}m_{\phi}^2\phi^2+\lambda_6 \phi^6,
\end{equation}
where the second term represents a tachyonic mass term, whose value is
assumed to be set by the soft SUSY breaking scale, $m_{\phi}\approx
m_{\rm soft}\approx 10^3{\rm GeV}$.  The energy scale of thermal
inflation is determined by the constant term $V_{\rm TI}$.  The exactly
flat potential is curved due to SUSY breaking, and stabilized by
unrenormalizable terms \footnote{ The exact form of the third term and
possible higher order terms are unimportant for our study.  }.  By
requiring the potential energy at the bottom of the potential to be
zero, we obtain $\lambda_6=\frac{m_{\phi}^6}{54V_{\rm TI}^2}$ and
$\phi_{\rm vev}=\sqrt{3V_{\rm TI}}/m_{\phi}$, where $\phi_{\rm vev}$ is
the vacuum expectation value of the flaton.

\begin{figure}[htbp]
 \begin{center}
  \includegraphics[width=170mm]{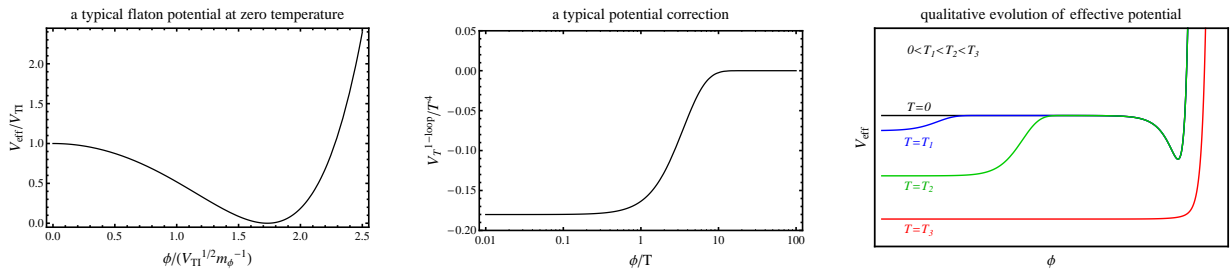}
 \end{center}
 \caption{The zero-temperature potential of the flaton and its finite-temperature correction.}
 \label{fig:one}
\end{figure}

Let us move on to the thermal corrections.  The one-loop effective
potential arising from thermal corrections is given by
\begin{equation}
V_T^{\rm 1-loop}(\phi)=T^4\sum_p g_p J_p\left(\frac{m_p(\phi,T)}{T}\right)\,,
\end{equation}
where $p$ labels both the bosonic and fermionic degrees of freedom and
the function $J_p$ is expressed in terms of an integral as
\begin{equation}
J_{\pm}(y)=\pm \frac{1}{2\pi^2}\int_0^{\infty}dx\,x^2 
{\rm ln}\left(1\mp e^{-\sqrt{x^2+y^2}}\right)\,,
\label{eq:def_J} 
\end{equation}
for bosons and fermions, respectively.  Following
Ref.\cite{Easther:2008sx}, the effective mass squared for fields in the
bath are
\begin{eqnarray}
\label{masssquareofothefields}
m_p^2(\phi,T)\approx \left\{ 
\begin{array}{lll}
m_{\rm b}^2+\frac{1}{2}\lambda_{\rm b}^2\phi^2+(\frac{1}{4}\lambda_{\rm b}^2+\frac{2}{3}g_{\rm b}^2)T^2 & {\rm boson\,,}\\
\frac{1}{2}\lambda_{\rm f}^2\phi^2 +\frac{1}{6}g_{\rm f}^2T^2 & {\rm fermion\,.}\\
\end{array}
\right. \,
\end{eqnarray}
Here we consider Yukawa couplings between the flaton and scalar
boson and fermion, with coupling constants $\lambda_{\rm b}$ and
$\lambda_{\rm f}$, respectively.  The coupling constants $g_{\rm b}$ and
$g_{\rm f}$ are associated with the gauge interactions of the scalar
boson and fermion, respectively.  We assume that the masses of other
bosons are also determined by $m_{\rm soft}\approx10^3$GeV and that
fermions are massless at tree level.  Since these corrections lower the
potential by ${\cal O}(T^4/10)$ around $|\phi| \lsim T$, there appears a
small dip at the origin, which traps the flaton to drive thermal
inflation.  We show an example flaton potential in Fig.\ref{fig:one}.

Thermal inflation begins when the energy density of other components
decays to be as small as the potential energy of the flaton.  If the
Universe is dominated by radiation, the temperature at the beginning
of thermal inflation, $T_{\rm begin}$, is given by $T_{\rm
begin}=\left(\frac{30}{\pi^2g_*}V_{\rm TI}\right)^{1/4}$.

During thermal inflation, the potential energy of the false vacuum phase
around the origin is larger than that of the true vacuum, meaning that
we might expect tunneling from the false to the true vacuum.  However,
the tunneling rate is so small \cite{Yamamoto:1985rd} that the flaton is
assumed to be fixed at the origin until the dip almost disappears.
Since the order of the curvature of the dip is determined by the
temperature as $V_{\rm eff}''\sim {\cal O}(T^2)$, thermal inflation ends
when the temperature becomes as small as $m_{\phi}\approx m_s$.
Therefore, by choosing $V_{\rm TI}$ and $m_{\phi}$, one can tune the
duration of thermal inflation.  Namely, the number of $e$-folds of thermal
inflation is roughly given by
\begin{equation}
{\cal N}=\log \left(\frac{T_{\rm begin}}{T_{\rm end}}\right)
\sim \log\left( \frac{V_{\rm TI}^{\frac{1}{4}}}{m_{\phi}} \right)\,.
\end{equation}
If we set $m_{\phi}=10^3$GeV and $V_{\rm TI}^{\frac{1}{4}}=10^7$GeV,
we obtain ${\cal N}\sim 9$.

Here we consider how thermal inflation solves the gravitino problem
\cite{gravitinoproblem}.  The gravitino, which only has suppressed
interactions and hence a long lifetime, decays after BBN and its decay
products affect the abundances of light elements.  As such, we can
constrain the abundance of the gravitino in the early Universe using
observations \cite{Kawasaki:2004qu}.  We use the variable
$Y_{3/2}=n_{3/2}/s$ to represent the comoving number density of
gravitinos, since the entropy density, $s$, is proportional to $a^{-3}$
if there is no entropy production.  Before the gravitinos decay, $Y_{3/2}$ is
approximately proportional to the reheating temperature $T_{\rm R}$.
Hence, if the reheating temperature is high, we have to decrease
$Y_{3/2}$.  According to Ref.\cite{Kawasaki:2004qu}, $T_{\rm
R}\gsim10^6\,{\rm GeV}$ may be problematic.  A solution proposed in
Ref.\cite{Lyth:1995hj} is to increase the entropy density via flaton decay after
thermal inflation.  The ratio of the entropy densities before and after
the flaton decay is
\begin{equation}
\label{entropyratio}
\frac{s_{\rm after}}{s_{\rm before}} \approx \frac{\frac{4}{3}\frac{V_{\rm TI}}{T_{\rm R,TI}}}{\frac{2\pi^2}{45}g_*(T_{\rm end})T_{\rm end}^3}
=1.5\times 10^{17}\left(\frac{V_{\rm TI}^{\frac{1}{4}}}{10^7\,{\rm GeV}} \right)^{4}  \left(\frac{T_{\rm R,TI}}{1\,{\rm GeV}} \right)^{-1}  
\left(\frac{T_{\rm end}}{1\,{\rm TeV}} \right)^{-3}  \left(\frac{g_*(T_{\rm end})}{200} \right)^{-1}\,,
\end{equation}
where $T_{\rm R,TI}$ is the reheating temperature associated with the
flaton decay.  Due to this significant entropy production the abundance
of gravitinos is made harmless.

As another possibility, let us consider the case where the Universe
transitions to thermal inflation after being dominated by oscillating
moduli.  Hereafter we use $\Phi$ to represent one of the moduli fields.
Since the moduli start oscillating when the Hubble parameter becomes as
small as the mass of the moduli ($m_{\Phi}$), they start oscillating
before reheating if the reheating temperature is lower than $\sim
\sqrt{m_{\Phi}M_{\rm Pl}}$. 
The energy density of the moduli at reheating is estimated as
\begin{equation}
\label{rhoPhi_at_reheating}
\rho_{\Phi}({\rm at \,\,reheating})=\frac{1}{2}m_{\Phi}^2\Phi_0^2 \times \left(\frac{a_{\rm osc}}{a_{\rm R}}\right)^3
=\frac{1}{2}m_{\Phi}^2\Phi_0^2 \times \left(\frac{H_{\rm R}}{H_{\rm osc}}\right)^2
=\frac{1}{2}\Phi_0^2 H_{\rm R}^2\,,
\end{equation}
where $\Phi_0$ is the initial amplitude of the oscillating moduli
and the subscript ``osc'' represents the value at the onset of oscillation. 
After reheating, since the temperature scales as $T\propto a^{-1}$,
$\rho_{\Phi}$ scales as $ \propto T^3$.
Therefore $T_{\rm begin}$ is determined by
\begin{equation}
\frac{1}{2}\Phi_0^2 H_{\rm R}^2 \times\left(\frac{T_{\rm begin}}{T_{\rm R}}\right)^3=V_{\rm TI}\,,
\end{equation}
then we obtain 
\begin{equation}
T_{\rm begin}\approx 9.7\times10^5 \left(\frac{g_*}{200}\right)^{-1/3}  \left(\frac{T_{\rm R}}{10^9\,{\rm GeV}}\right)^{1/3}  
\left(\frac{V_{\rm TI}^{\frac{1}{4}}}{10^7\,{\rm GeV}}\right)^{4/3}  \left(\frac{\Phi_0}{M_{\rm Pl}}\right)^{1/3}\, {\rm GeV}\,.
\end{equation}
On the other hand, if the reheating temperature is high, the oscillations
begin in the radiation-dominated Universe, when
\begin{equation}
\label{ToscinB}
m_{\Phi}^2=H_{\rm osc}^2=\frac{\pi^2}{90}g_*\frac{T_{\rm osc}^4}{M_{\rm Pl}^2}\,,
\end{equation}
is satisfied.
If $\Phi_0$ is as large as $M_{\rm Pl}$,
the energy density
associated with the coherent oscillation of moduli soon becomes dominant.
In this case $T_{\rm begin}$
is determined by
\begin{equation}
\frac{1}{2}m_{\Phi}^2 \Phi_0^2\times\left(\frac{T_{\rm begin}}{T_{\rm osc}}\right)^3=V_{\rm TI}\,,
\end{equation}
combining the above expressions we obtain
\begin{equation}
T_{\rm begin}\approx 3.4\times 10^5 \left(\frac{g_*}{200}\right)^{-1/4} \left(\frac{V_{\rm TI}^{\frac{1}{4}}}{10^7\,{\rm GeV}}\right)^{4/3}
 \left(\frac{m_{\Phi}}{1\,{\rm TeV}}\right)^{-1/6}  \left(\frac{\Phi_0}{M_{\rm Pl}}\right)^{2/3}\,{\rm GeV}\,.
\end{equation}

Let us move on to the cosmological moduli problem.  The moduli abundance
$Y_{\Phi}=n_{\Phi}/s$ should also be small enough so as not to spoil BBN
\cite{Ellis:1990nb}.  Assuming the moduli start oscillating before
reheating, during the era when the energy density associated with the
coherent oscillations of the inflaton dominate the Universe, the moduli
abundance before flaton decay is evaluated as
\begin{equation}
Y_{\Phi}=\frac{\frac{1}{m_{\Phi}}\frac{1}{2}\Phi_0^2H_{\rm R}^2}{\frac{4}{3T_{\rm R}}\times 3M_{\rm Pl}^2H_{\rm R}^2}
=\frac{1}{8}\frac{T_{\rm R}}{m_{\Phi}} \left(\frac{\Phi_0}{M_{\rm Pl}}\right)^2\,,
\end{equation}
where we use eq. (\ref{rhoPhi_at_reheating})
and assume that there is no entropy production after reheating.
After the flaton decays, by using eq. (\ref{entropyratio}), $Y_{\Phi}$ becomes
\begin{align}
Y_{\Phi,\,{\rm after}}\approx& \,\, \frac{\pi^2}{240}g_*(T_{\rm end}) \left(\frac{\Phi_0}{M_{\rm Pl}}\right)^2 \frac{T_{\rm R}T_{\rm R,TI}T_{\rm end}^3}{m_{\Phi}V_{\rm TI}}\,\notag\\
=&\,\,8.2\times10^{-13}   \left(\frac{V_{\rm TI}^{\frac{1}{4}}}{10^7\,{\rm GeV}}\right)^{-4}  
 \left(\frac{T_{\rm R}}{10^9\, {\rm GeV}}\right)
   \left(\frac{T_{\rm R,TI}}{1\,{\rm GeV}}\right)   \left(\frac{T_{\rm end}}{m_{\Phi}}\right)^{3}  \left(\frac{m_{\Phi}}{1\,{\rm TeV}}\right)
    \left(\frac{\Phi_0}{M_{\rm Pl}}\right)^2 \left(\frac{g_*(T_{\rm end})}{200}\right)  \,.
\end{align}
Therefore, with appropriate parameters, thermal inflation can make
$Y_{\rm \Phi}$ small enough for successful BBN.

\section{flaton dynamics in a thermal bath \label{fdtb}}
In this section, we consider the flaton dynamics based on
finite-temperature field theory.  In order to describe the dynamics of
the expectation values of quantum fields in a thermal bath, we use the
effective action method, which has been studied in several contexts
\cite{FTEA,Yamaguchi:1996dp,Yokoyama:2004pf,Greiner:1996dx} based on the in-in or
the closed time-path formalisms.  Using this method, we can evaluate the
evolution of expectation values by performing path integrals along two
time paths, with two field variables $\phi_{\pm}$ defined on each path.
Generally the effective action can be expressed as
\cite{FTEA,Yamaguchi:1996dp,Yokoyama:2004pf,Greiner:1996dx}
\begin{equation}
\Gamma= S_0+ \Gamma_{\rm R} + \Gamma_{\rm I} \,,
\end{equation}
where $S_0$ is the tree level action, and $\Gamma_{\rm R}$ and $\Gamma_{\rm
I}$, respectively, represent the real and imaginary parts coming from
interactions.  The imaginary part has the following structure
\begin{align}
\exp \left[i\Gamma_{\rm I}\right]
&= \exp\left[-\frac{1}{2}\int d^4x_1 d^4x_2 A_{\rm a}(x_1-x_2) \phi_{\Delta}(x_1) \phi_{\Delta}(x_2) 
+A_{\rm m}(x_1-x_2)\phi_{\Delta}(x_1)\phi_{\Delta}(x_2) \phi_c(x_1)\phi_c(x_2) 
\right]\,,
\end{align}
and we can rewrite it as
\begin{align}
\exp \left[i\Gamma_{\rm I}\right]=\int{\cal D}\xi_{\rm a}
{\cal D}\xi_{\rm m} 
\, P[\xi_{\rm a}] P[\xi_{\rm m}] 
\exp\left[i S_{\rm noise}\right]
\,,
\end{align}
where
\begin{align}
P[\xi_{\rm a}]&\propto \exp\left[-\frac{1}{2}\int d^4x_1 d^4x_2\, \xi_{\rm a}(x_1)A_{\rm a}^{-1}(x_1-x_2)\xi_{\rm a}(x_2) \right] \,,\notag\\
P[\xi_{\rm m}]&\propto \exp\left[-\frac{1}{2}\int d^4x_1 d^4x_2\, \xi_{\rm m}(x_1)A_{\rm m}^{-1}(x_1-x_2)\xi_{\rm m}(x_2) \right] \,,\notag\\
S_{\rm noise}&=\int d^4x\, \left[\xi_{\rm a}(x)\phi_{\Delta}(x) +\xi_{\rm m}(x)\phi_{\Delta}(x)\phi_{c}(x) \right]\,, \notag\\
&\phi_{c}=\frac{\phi_++\phi_-}{2}\,,  \quad \phi_{\Delta}=\phi_+-\phi_-\,.
\end{align}
We can interpret the new variables $\xi_{\rm a}$ and $\xi_{\rm m}$ as
stochastic noises whose probability distributions are given by
$P[\xi_{\rm a}]$ and $P[\xi_{\rm m}]$, respectively.  Finally the
equation of motion for $\phi_c$, which is obtained by
varying the effective action with respect to $\phi_{\Delta}$, becomes the
Langevin equation
\begin{align}
&\Box \phi(x)+V_{\rm eff}'[\phi]
+\int_{-\infty}^t dt'\int d^3x'\,B_{\rm a}(x-x')\phi(x')
+\phi(x)\int_{-\infty}^tdt'\int d^3x'\,B_{\rm m}(x-x')\phi^2(x')\notag\\
=&\xi_{\rm a}(x)+\xi_{\rm m}(x)\phi(x)\,.
\end{align}
We briefly see specific examples studied in Ref. \cite{Yokoyama:2004pf}.
An interaction term ${\cal L}_{\rm int}=-\lambda^2 \chi^2\phi^2$, where $\chi$ is a real scalar field,
 leads to both additive and multiplicative noises
and the corresponding non-local terms.
Functions $A$ and $B$ for additive noise and the correspondent non-local terms in Fourier space are calculated as
\begin{align}
A_{\rm a}(\omega, \vec{k})=-16\pi i \lambda^4
\int  &\frac{d^3q_1}{(2\pi)^3}\frac{d^3q_2}{(2\pi)^3}\frac{d^3q_3}{(2\pi)^3} (2\pi)^3 \delta^3(\vec{q}_1+\vec{q}_2+\vec{q}_3-\vec{k})
\frac{1}{8\omega_{q_1}\omega_{q_2}\omega_{k-q_1-q_2}}\notag\\
\times \Large[&\left\{(1+n_{q_1})(1+n_{q_2})(1+n_{q_3})+n_{q_1} n_{q_2}n_{q_3}\right\}\delta(\omega-\omega_{q_1}-\omega_{q_2}-\omega_{q_3})\notag\\
+&\left\{(1+n_{q_1})(1+n_{q_2})n_{q_3}+n_{q_1} n_{q_2}(1+n_{q_3})\right\}\delta(\omega-\omega_{q_1}-\omega_{q_2}+\omega_{q_3})\notag\\
+&\left\{(1+n_{q_1})n_{q_2}(1+n_{q_3})+n_{q_1} (1+n_{q_2})n_{q_3}\right\}\delta(\omega-\omega_{q_1}+\omega_{q_2}-\omega_{q_3})\notag\\
+&\left\{n_{q_1}(1+n_{q_2})(1+n_{q_3})+(1+n_{q_1}) n_{q_2}n_{q_3}\right\}\delta(\omega+\omega_{q_1}-\omega_{q_2}-\omega_{q_3})\notag\\
+&\left\{(1+n_{q_1})n_{q_2}n_{q_3}+n_{q_1} (1+n_{q_2})(1+n_{q_3})\right\}\delta(\omega-\omega_{q_1}+\omega_{q_2}+\omega_{q_3})\notag\\
+&\left\{n_{q_1}(1+n_{q_2})n_{q_3}+(1+n_{q_1}) n_{q_2}(1+n_{q_3})\right\}\delta(\omega+\omega_{q_1}-\omega_{q_2}+\omega_{q_3})\notag\\
+&\left\{n_{q_1}n_{q_2}(1+n_{q_3})+(1+n_{q_1})(1+ n_{q_2})n_{q_3}\right\}\delta(\omega+\omega_{q_1}-\omega_{q_2}+\omega_{q_3})\notag\\
+&\left\{n_{q_1} n_{q_2}n_{q_3}+(1+n_{q_1})(1+n_{q_2})(1+n_{q_3})\right\}\delta(\omega+\omega_{q_1}+\omega_{q_2}+\omega_{q_3})\Large]\,,
\end{align}
\begin{align}
B_{\rm a}(\omega, \vec{k})= 8\pi \lambda^4
\int  &\frac{d^3q_1}{(2\pi)^3}\frac{d^3q_2}{(2\pi)^3}\frac{d^3q_3}{(2\pi)^3} (2\pi)^3 \delta^3(\vec{q}_1+\vec{q}_2+\vec{q}_3-\vec{k})
\frac{1}{8\omega_{q_1}\omega_{q_2}\omega_{q_3}}\notag\\
\times\Large[&\left\{(1+n_{q_1})(1+n_{q_2})(1+n_{q_3})-n_{q_1} n_{q_2}n_{q_3}\right\}\delta(\omega-\omega_{q_1}-\omega_{q_2}-\omega_{q_3})\notag\\
+&\left\{(1+n_{q_1})(1+n_{q_2})n_{q_3}-n_{q_1} n_{q_2}(1+n_{q_3})\right\}\delta(\omega-\omega_{q_1}-\omega_{q_2}+\omega_{q_3})\notag\\
+&\left\{(1+n_{q_1})n_{q_2}(1+n_{q_3})-n_{q_1} (1+n_{q_2})n_{q_3}\right\}\delta(\omega-\omega_{q_1}+\omega_{q_2}-\omega_{q_3})\notag\\
+&\left\{n_{q_1}(1+n_{q_2})(1+n_{q_3})-(1+n_{q_1}) n_{q_2}n_{q_3}\right\}\delta(\omega+\omega_{q_1}-\omega_{q_2}-\omega_{q_3})\notag\\
+&\left\{(1+n_{q_1})n_{q_2}n_{q_3}-n_{q_1} (1+n_{q_2})(1+n_{q_3})\right\}\delta(\omega-\omega_{q_1}+\omega_{q_2}+\omega_{q_3})\notag\\
+&\left\{n_{q_1}(1+n_{q_2})n_{q_3}-(1+n_{q_1}) n_{q_2}(1+n_{q_3})\right\}\delta(\omega+\omega_{q_1}-\omega_{q_2}+\omega_{q_3})\notag\\
+&\left\{n_{q_1}n_{q_2}(1+n_{q_3})-(1+n_{q_1})(1+ n_{q_2})n_{q_3}\right\}\delta(\omega+\omega_{q_1}-\omega_{q_2}+\omega_{q_3})\notag\\
+&\left\{n_{q_1} n_{q_2}n_{q_3}-(1+n_{q_1})(1+n_{q_2})(1+n_{q_3})\right\}\delta(\omega+\omega_{q_1}+\omega_{q_2}+\omega_{q_3})\Large]\,,
\end{align}
where $\omega_{q_i}=\sqrt{|\vec{q}_i|^2 +m_{\chi}^2}$ and $n_{q_i}=\frac{1}{e^{\beta \omega_{q_i}}-1}$.
For the multiplicative noise and corresponding non-local term, we find
\begin{align}
A_{\rm m}(\omega, \vec{k})=2\pi \lambda^4
\int \frac{d^3q}{(2\pi)^3} \frac{1}{\omega_{q}\omega_{k-q}}
\times \Large[&\left\{(1+n_{q})(1+n_{k-q})+n_q n_{k-q}\right\}\delta(\omega-\omega_q-\omega_{k-q})\notag\\
+&\left\{(1+n_{q})n_{k-q}+n_q (1+n_{k-q})\right\}\delta(\omega-\omega_q+\omega_{k-q})\notag\\
+&\left\{(n_{q}(1+n_{k-q})+(1+n_q) n_{k-q}\right\}\delta(\omega+\omega_q-\omega_{k-q})\notag\\
+&\left\{n_q n_{k-q}+(1+n_{q})(1+n_{k-q})\right\}\delta(\omega+\omega_q+\omega_{k-q})\Large]\,,
\end{align}
\begin{align}
B_{\rm m}(\omega, \vec{k})=-2\pi i \lambda^4
\int \frac{d^3q}{(2\pi)^3} \frac{1}{\omega_{q}\omega_{k-q}}
\times\Large[&\left\{(1+n_{q})(1+n_{k-q})-n_q n_{k-q}\right\}\delta(\omega-\omega_q-\omega_{k-q})\notag\\
+&\left\{(1+n_{q})n_{k-q}-n_q (1+n_{k-q})\right\}\delta(\omega-\omega_q+\omega_{k-q})\notag\\
+&\left\{(n_{q}(1+n_{k-q})-(1+n_q) n_{k-q}\right\}\delta(\omega+\omega_q-\omega_{k-q})\notag\\
+&\left\{n_q n_{k-q}-(1+n_{q})(1+n_{k-q})\right\}\delta(\omega+\omega_q+\omega_{k-q})\Large]\,.
\end{align}
Though the noise terms generally consist of both additive noise, $\xi_{\rm a}$, 
and multiplicative noise, $\xi_{\rm m}\phi$, we focus on the
additive noise term since the former is more important to trigger phase transition. 
This noise term is related to the
``friction'' term
through the fluctuation-dissipation relation \cite{Yokoyama:2004pf,Greiner:1996dx}
\begin{equation}
\frac{\rm noise \,\,correlation}{\rm dissipation\,\, coefficient}
=\frac{A_{\rm a}(\omega,\vec{k})}{iB_{\rm a}(\omega,\vec{k})/2\omega}
=\omega \frac{e^{\omega/T}+1}{e^{\omega/T}-1}
\to 2T \,\,(T\gg \omega)\,.
\end{equation}

In Ref. \cite{Yamaguchi:1996dp} it was shown that the damping scale of
the fermionic noise correlation is independent of the mass
of the fermion, which is
different from the bosonic noise whose correlation damps exponentially
above the mass scale.  Therefore, in the high-temperature regime $T\gg
m$, the dominant noise component comes from interactions with fermions.
More quantitatively, the correlation function for fermionic noise can be
expressed as
\begin{equation}
\langle \xi(t,\vec{x})\xi(t,\vec{x}')\rangle\propto \frac{T^4}{r^2}e^{-2\pi rT}\,, \quad {\rm for}\,\,\, r\gg\frac{1}{\pi T}\,,\quad(r=|\vec{x}-\vec{x}'|)\,.
\end{equation}
From this expression we take the correlation length of thermal noise as
$(\pi T)^{-1}$.  This length scale is very important in estimating the
typical value of the flaton at finite temperature.  Here let us take a
quick look at this typical field value, as this will help us to
understand the results of numerical simulations later.  The form of the
effective potential is too complicated to be well approximated by a
simple polynomial function, so for simplicity let us neglect the
potential here.  Following Ref. \cite{Yamaguchi:1997sy}, the mean square
value of the coarse-grained field $\phi$ over the spatial scale $R$ is given
by
\begin{equation}
\label{phito2R}
\langle \phi^2 \rangle_{\rm R} =\frac{1}{2\pi^2}\int_0^{\infty}dk \,k\left(\frac{1}{2} +\frac{1}{e^{\frac{k}{T}}-1}\right)\,W(k,R)^2\,,
\end{equation}
where $W(k,R)$ is the coarse-graining window function.
As an example, if we take the Gaussian function
\begin{equation}
W(k,R)=e^{-\frac{1}{2}k^2R^2}\,,
\end{equation}
we obtain $\sqrt{\langle \phi^2 \rangle}\approx 0.43T$ for $R=(\pi T)^{-1}$.

Since the correlation length of the noise is $\sim (\pi T)^{-1}$, we can
treat the noise as being uncorrelated on larger scales.  The same is
true for the temporal noise correlation, since it is suppressed
exponentially for $\Delta t > (\pi T)^{-1}$.  As such, the noise term
can be approximated by a white, Gaussian random variable when we
consider dynamics on spatial and temporal scales that are larger
than the above correlation length.  Hence we use the following simple
EoM.
\begin{equation}
\label{simpleEOM}
\ddot \phi(\vec{x},t) -\vec{\nabla}^2\phi(\vec{x},t) +\eta \dot\phi(\vec{x},t) +V_{\rm eff}'[\phi]=\xi(\vec{x},t)\,,
\end{equation}
where the correlation function of the noise term is
\begin{equation}
\left\langle \xi(\vec{x},t) \xi(\vec{x}',t')\right\rangle=D\delta(t-t')\delta^3(\vec{x}-\vec{x}')\,.
\end{equation}
The fluctuation-dissipation relation in this simple EoM is
\begin{equation}
\frac{D}{\eta}=2T\,.
\end{equation}
Due to the fluctuation-dissipation relation, equilibrium values do not
depend on the friction coefficient $\eta$.  
Its value is related with the decay rate of $\phi$ particle
if $\phi$ is oscillating \cite{FTEA,Yokoyama:2004pf,Greiner:1996dx}.
On dimensional grounds we can take $\Gamma \propto T$. 
Since the value of $\eta$ only determines
the time scale on which the system approaches equilibrium,
here we simply take $\eta =T$ as strong enough couplings between the flaton and the thermal bath are required 
for successful thermal inflation.
Then the ratio of the equilibration
timescale to the cosmic expansion timescale is
\begin{equation}
\frac{\rm equilibration\,\, timescale}{\rm Hubble\,\, time}\sim
\frac{\eta^{-1}}{H^{-1}} =
\frac{T^{-1}}{H^{-1}} \sim
\begin{cases}
    \frac{T}{M_{\rm Pl}} & (\rm RD \,\,era) \,,\\
  &\\
    \frac{V_{\rm TI}^{\frac{1}{2}}}{M_{\rm Pl}\,T}  & ( \rm during\,\, thermal\,\, inflation)\,.
  \end{cases}
\end{equation}
We see that this ratio is much smaller than unity in both the RD era and
the period of thermal inflation,
from which we can conclude that the equilibration time is still much shorter than
the Hubble time even if we take other choices for the value of $\eta$.
This huge difference between the two timescales allows us to safely ignore the
Hubble expansion in simulations we show later.

\section{Setup of Numerical Simulations \label{simulationsetup}}

In this section we summarize the details of our three-dimensional
lattice simulation.  We solved the equation of motion given by
eq. (\ref{simpleEOM}) by the second-order explicit Runge-Kutta method
with the second-order finite differences approximating the spatial
derivatives.  The basic setup is the same as in
Ref.\cite{Yamaguchi:1996dp}.  In numerical calculations we use
dimensionless variables like $\tilde x=Tx$, $\tilde t=Tt$, $\tilde
\phi=\phi/T$, and $\tilde \xi=\xi/T^3$ since the scale of interest is
deeply related to the temperature.

The noise correlation function on the lattice becomes
\begin{equation}
\langle \xi(\vec{x}_i,t_m)\xi(\vec{x}_j,t_n)\rangle=2\eta \delta(t_m-t_n)\delta^3(\vec{x}_i-\vec{x}_j)
\to \frac{2\eta}{\Delta t (\Delta x)^3} \delta_{m,n} \delta_{i,j}\,,
\end{equation}
since on the lattice the delta functions are properly replaced as
$\delta(t_m-t_n) \to (\Delta t)^{-1}\delta_{m,n}$ and $
\delta^3(\vec{x}_i-\vec{x}_j)\to (\Delta x)^{-3}\delta_{i,j}$.  The
value of noise variable on each lattice is given by
\begin{equation}
\xi(\vec{x}_i,t_m)=\left(\frac{2\eta}{\Delta t (\Delta x)^3}\right)^{\frac{1}{2}} {\mathcal G}_{i,m}\,,
\end{equation}
where ${\mathcal G}$ is a standard Gaussian random variable.

We also define approximation function of the potential term, which is
shown in Appendix.  As can be seen later, the quantitative shape of the
effective potential is very sensitive to the temperature, especially at
the end of thermal inflation.  Therefore we use the above approximation
function both in the lattice simulation and semi-analytic calculation.

We choose the initial condition for simulations as
\begin{equation}
\phi(\vec{x},t=0)=\dot\phi(\vec{x},t=0)=0 \,.
\end{equation}
Although this is an admittedly unrealistic initial condition, we have
confirmed that the field quickly reaches the thermal configuration
compared to the typical duration of simulation time and the timescale of
the temperature variation.

With the above settings we use the $256^3$ lattice points and $m_{\phi}$
(and $m_{\rm b}$ in eq. (\ref{masssquareofothefields}))$=10^3$ and
$10^2$ GeV, but the qualitative results do not depend on these mass
values.

\section{Results of Numerical Simulations \label{RNS}}
\subsection{phase 1: before thermal inflation}

A necessary initial condition for the flaton to drive thermal inflation
is that the field value of the flaton should be homogeneously close to
zero before thermal inflation begins.  However, the form of the 1-loop
effective potential suggests that there is more than one local minimum,
and if the flaton field is trapped in the true vacuum in some spatial
regions, the thermal inflation scenario does not work.  In order to
determine whether or not this problem is encountered, we simulated the
time evolution of the flaton from a very high temperature, $T_{\rm 0}$,
to the temperature at which thermal inflation begins.

The ``high'' temperature $T_{\rm 0}$ is determined by the following
consideration.  In order to realize a situation where the typical value
of the flaton is $\phi_{\rm vev}$ ($\equiv \sqrt{3V_{\rm TI}}/m_{\phi}$,
the vacuum expectation value at $T=0$), we first perform a simulation at
$T=\phi_{\rm vev}$, expecting $\sqrt{\langle \phi^2 \rangle}\approx
T\approx \phi_{\rm vev}$.\footnote{ Note that the VEV of the
zero-temperature potential also depends on $V_{\rm TI}$ as $\phi_{\rm
vev}=\sqrt{3V_{\rm TI}}/m_{\phi}$.  Since the temperature at the
beginning of thermal inflation, $T_{\rm begin}$, is controlled by
$V_{\rm TI}$ (see Section \ref{TIS}), we choose the value of $V_{\rm
TI}$ such that the number of $e$-folds of thermal inflation becomes
about $6$.  In order to calculate the number of e-folds we also need to
know the temperature at the end of thermal inflation, and this can be
determined once we have fixed the coupling constants.}  At this
temperature the shape of the effective potential becomes like the
potential labelled ``$T=T_1$'' in the right panel of Fig.\ref{fig:one}.
We then perform a second simulation, setting the temperature to half of
that in the previous simulation and using the final configuration of the
previous simulation to determine the initial conditions.  Since we fix
the gridsize of the simulation and the value of the lattice spacing
normalized by the temperature, the physical size of the second
simulation box is larger than that of the previous, hotter simulation.
We therefore use periodic boundary
conditions and define the initial condition for $\phi$ and $\dot \phi$
as averaged quantities of the previous values of close grids on each new
grids.  Repeating this procedures $N$ times we can follow the flaton
dynamics from $T=T_0$ to $T = T_0 \times 2^{-N}\sim T_{\rm begin}$.

\begin{figure}[htbp]
 \begin{center}
  \includegraphics[width=90mm]{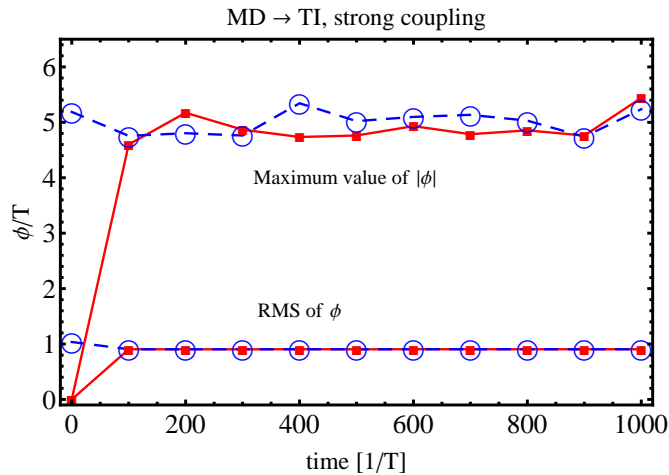}
 \end{center}
   \begin{minipage}{0.95\hsize}
  \begin{center}
 \caption{The results of one example multistage lattice simulation that
 was performed assuming moduli-domination before thermal inflation and
 strong coupling to the fields in the thermal bath.  In other cases the
 results are qualitatively the same.  The root mean square of $\phi$ and
 the maximum value of $|\phi|$ in the first and the last simulation at
 each reference time are shown.  The red lines with square vertices are
 the results of the first (hot) simulation and the dashed blue lines
 with circular vertices are those of the last simulation, where $T\sim
 T_{\rm begin}$.  Since we impose the initial conditions $\phi=\dot
 \phi=0$ in the first simulation and the initial conditions for the
 following simulations are determined sequentially by the final
 configuration of the preceding, higher temperature simulation, the
 flaton distribution at each first reference time is not the equilibrium
 configuration.
  \label{multistage_figure}}
  \end{center}
  \end{minipage}
\end{figure}

In the numerical simulations we consider corrections to the potential
coming from a single bosonic and single fermionic degree of freedom.  In
order to try and establish the importance of the thermal effects we
perform simulations with two choices of the coupling constants appearing
in eq. (\ref{masssquareofothefields}).  Hereafter we refer to these two
choices as the strongly and weakly coupled cases, and they correspond to
taking $\lambda_{\rm b}=g_{\rm b}=\lambda_{\rm f}=g_{\rm f}=1$ and
$\lambda_{\rm b}=g_{\rm b}=\lambda_{\rm f}=g_{\rm f}=0.1$ respectively.
We also consider two different scenarios.  In the first scenario thermal
inflation is preceded by moduli domination (MD$\to$TI) and in the second
scenario thermal inflation is preceded by radiation domination
(RD$\to$TI).  The results of one example simulation are shown in
Fig.\ref{multistage_figure}.  For the form of effective potential used in this
study, we confirm that the typical value of the flaton is $\sqrt{\langle
\phi^2 \rangle}\approx T$, regardless of the temperature before thermal
inflation.  In other words, we do not see any spatial regions where the
field value remains so large that the flaton potential energy becomes
inhomogeneous and ruins the thermal inflation scenario.

We close this subsection with comments on the validity of our multistage simulation.
The result shown in Fig. \ref{multistage_figure} confirms us that 
we properly follow the dynamics of the flaton from a high temperature to
$T_{\rm begin}$, with multistage simulation.
Since the equilibration timescale ($\sim \eta^{-1}$) is much shorter than that of temperature change ($\sim H^{-1}$),
the system approaches the equilibrium rapidly enough in each simulation with a fixed temperature.
In other words, even though we impose out-of-equilibrium initial condition 
which is simply connected by the previous simulation where the temperature is set twice as hot,
we can realize the equilibrium distribution ($\sqrt{\langle \phi^2 \rangle} \sim T$) by performing a simulation 
for a longer time than $\eta^{-1}$ (but much shorter than $H^{-1}$).
Therefore repetitive simulations enable us to consider a system in
quasi-equilibrium state for a longer time than Hubble time
without including the exact change in temperature.
The smooth change of the root mean square (RMS) value
obtained in Fig. \ref{multistage_figure} justifies a factor of 2 change of the temperature at each step
is small enough to warrant the adiabatic change of the temperature in the sequential simulations.
As for the maximum value, we note that for random $256^3$ realization of Gaussian distribution, 
the probability the maximun exceeds 6.2$\sigma$ (5.6T) is 1 $\%$ and that it lies lower than 5.2 $\sigma$ (4.6T) is also 1$\%$.
Although the field value at each point is correlated with nearby points,
we find one-point distribution function is close to a Gaussian distribution.
Hence we may conclude the observed
maximum values in Fig. \ref{multistage_figure} are also in accordance with the entire distribution.

\begin{table}[htbp]
 \begin{center}
  \begin{tabular}{|c|c|c|c|} \hline \rule[0pt]{0pt}{8pt}
  scenario & 
  couplings &
  $T_{\rm begin}$[GeV] & 
  $\phi_{\rm vev}(=T_0)$[GeV]  \\ \hline \rule[0pt]{0pt}{10pt}
   MD $\to$ TI &strong & $2.1\times10^6$ & $3.7\times 10^{12}$ \\\hline \rule[0pt]{0pt}{10pt}
   MD $\to$ TI &weak & $1.7\times 10^7$ & $ 8.3\times 10^{13}$\\\hline \rule[0pt]{0pt}{10pt}
   RD $\to$ TI &strong & $2.1\times10^6$ & $6.0\times10^{10}$ \\\hline \rule[0pt]{0pt}{10pt}
   RD $\to$ TI &weak & $1.7\times 10^7$ & $3.9\times 10^{12}$ \\\hline
  \end{tabular}
  \caption{The temperature at the beginning of thermal inflation and VEV
  of the flaton.  Since the ratio of these values is ${\cal
  O}(10^6)\sim 2^{20}$, we performed about 20 simulations to follow the
  flaton dynamics from $T_0$ to $T_{\rm begin}$.  }
   \end{center}
  \label{phase1table}
\end{table}

\subsection{phase 2: at the end of thermal inflation}

It is believed that thermal inflation ends with a first-order phase
transition accompanied by the formation of bubbles, and that the
collision of these bubbles then leads to gravitational wave production.
Here we briefly review the theory of tunneling at a finite temperature
and define the percolation temperature at which the bubbles collide and
start generating gravitational waves.

The tunneling rate per unit volume at temperature $T$
is estimated as \cite{Linde:1980tt}
\begin{equation}
\label{tunnelingrateFT}
\Gamma(T) \sim T^4 e^{-\frac{S_3}{T}}\,,
\end{equation}
where $S_3$ is the Euclidean action after performing the time integral,
\begin{equation}
S_3=\int d^3x \left(\frac{1}{2}\left(\nabla \phi \right)^2 +V(\phi) \right)\,.
\end{equation}
The dominant contribution to the tunneling rate comes from the solution
of the equation of motion,
\begin{equation}
\frac{d^2\phi}{dr^2}+\frac{2}{r}\frac{d\phi}{dr}-\frac{dV}{d\phi}=0\,,\quad(r=|\vec{x}|)
\end{equation}
under the boundary conditions
$\phi(r=\infty)=0$ and $\left.\frac{d\phi}{dr}\right|_{r=0}=0$.

The fraction of spatial regions occupied by bubbles can be written
as \cite{Guth:1981uk}
\begin{equation}
F(t)=1-e^{-P(t)}\,,
\end{equation}
where the function $P(t)$ is given by
\begin{align}
P(t)=&\int ^t dt'\, \Gamma(t') \frac{4\pi}{3}\left( \int _{t'}^t dt''\, \frac{a(t)}{a(t'')}\right)^3 \notag\\
=&\frac{4\pi}{3}\int ^t dt'\, \Gamma(t')\frac{1}{H^3}\left( e^{H(t-t')}-1 \right)^3\,.
\end{align}
Making use of eq.(\ref{tunnelingrateFT}) we can rewrite this in terms of
temperature as
\begin{equation}
P(T)=\frac{4\pi}{3} \int_{T}^{\infty} dT' \,\frac{T'^3}{H^4} \left(\frac{T'}{T}-1\right)^3 e^{-\frac{S_3(T)}{T}} \,.
\end{equation}
In this paper we define the percolation temperature as $F(T=T_{\rm
p})=0.5$ \footnote{ The qualitative conclusion ($T_{\rm curv}\approx
T_{\rm p}<T_{\rm sub}$) remains unchanged if we employ other definitions
such as $F(T_{\rm p})=0.01$ or $0.99$.  } .  Note that since the
exponential factor $\exp[-S_3(T)/T]$ is very sensitive to the
temperature and quickly becomes small when we take a large value of $T$,
it is sufficient to take the upper limit of the integral to be some
finite value.  For example, it is enough to take it as $2T_{\rm curv}$,
where $T_{\rm curv}$ is the temperature at which the curvature of the
potential becomes zero.  After evaluating the above quantities
numerically, we find that the difference between the percolation
temperature $T_{\rm p}$ and $T_{\rm curv}$ is tiny, so that the Universe
becomes filled with critical bubbles almost immediately after bubble
formation effectively begins.

From the above consideration based on the shape of the flaton effective
potential, we may expect that thermal inflation ends with a
first-order phase transition characterized by critical bubble formation.
However, this description is based on the assumption that the flaton is
well within the false vacuum phase before bubble nucleation occurs.

\begin{figure}[htbp]
 \begin{center}
  \includegraphics[width=170mm]{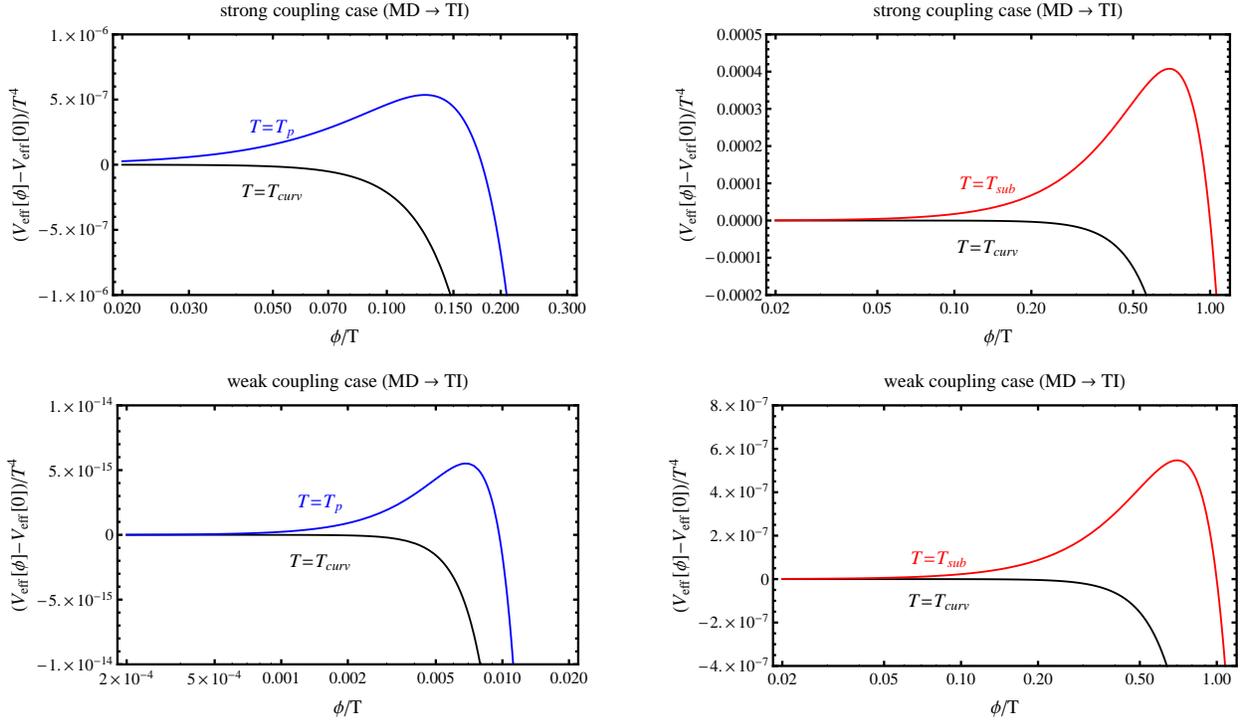}
 \end{center}
 \caption{Some examples of the effective potential at $T=T_{\rm curv}$,
 $T_{\rm p}$, and $T_{\rm sub}$ are shown.  Since at $T=T_{\rm p}$ the
 local maximum is located at $\phi<T$ and its height is much smaller
 than $T^4$, the flaton is able to escape the local minimum and critical
 bubble formation theory is not applicable.  } \label{finalpotential}
\end{figure}

We see from Fig.\ref{finalpotential} that around the percolation
temperature the potential barrier is located at $\phi \ll T$ and the
height of the barrier is much smaller than $T^4$.  Taking thermal
fluctuations into account, since the width of the field distribution is
$\sqrt{\langle \phi^2 \rangle}\approx T$, we conclude that the small
potential barrier cannot trap the flaton in the false vacuum phase until
the temperature becomes as small as the temperature at which critical
bubble nucleation occurs.  
This means that the two phases coexist well before the percolation epoch
in the bubble nucleation picture, and the phase transition proceeds with
phase-mixing.
As such, the standard description of the end of thermal
inflation in terms of a strong first-order phase transition which is
accompanied with bubble formation is inappropriate.

Now let us investigate more quantitatively the failure of critical
bubble formation as a description of the end of thermal inflation.  The
width of the wall trapping the flaton is broad at high temperatures and
gradually becomes thin as the temperature drops.  We define the width
in field space, $\phi_{\rm wid}$, at temperature $T$, as
\begin{equation}
V_{\rm eff}[\phi=\phi_{\rm wid},T]=V_{\rm eff}[\phi=0,T]\,.
\end{equation}
Since the shape of the effective potential depends on temperature, we
obtain $\phi_{\rm wid}(T)$ by solving the above equation.  As a typical
temperature at which phase-mixing occurs, we define the temperature
$T_{\rm sub}$ as
\begin{equation}
\phi_{\rm wid}(T=T_{\rm sub})=T_{\rm sub}\,,
\end{equation}
i.e. $T_{\rm sub}$ is the temperature at which the width of the
potential wall becomes as small as the temperature.  As we see from the
simulations in the previous subsections and the analytical estimation
(eq. (\ref{phito2R})), the typical value of $\phi$ is as large as $T$.
Therefore, at $T=T_{\rm sub}$, and if the height of the potential
barrier is small enough, spatial regions in which the flaton lies
outside of the potential dip are ubiquitous in the Universe.  We call
such regions subcritical bubbles \cite{subcriticalbubble}, 
which are continuously created and destroyed 
by thermal fluctuations and hence differ from the critical bubbles 
which only grow after being nucleated by tunneling.
For the effective potential we study in this paper, the
relations $T_{\rm sub}>T_{\rm p}$ and $F(T_{\rm sub})\ll1$ hold.
Therefore, at $T=T_{\rm sub}$ the flaton is no longer trapped at the
local minimum at the origin, meaning that there are practically no
critical bubbles.  Specific values are shown in Table \ref{temptable}.
We would like to make a comment on the temperature at
the end of thermal inflation, $T_{\rm end}$ quantitatively. 
In Section \ref{TIS} we estimated $T_{\rm end} \sim m_{\phi}$.
Table \ref{temptable}, however, shows that while $T_{\rm curv}$, $T_{\rm p}$, and $T_{\rm sub}$ coincide with each other
within $5 \%$ they deviate from $m_{\phi}$ by a factor of 5 - 40.
Hence we should use $T_{\rm end}\sim T_{\rm sub}$ to estimate the proper duration of thermal inflation.

\begin{figure}[htbp]
 \begin{center}
  \includegraphics[width=90mm]{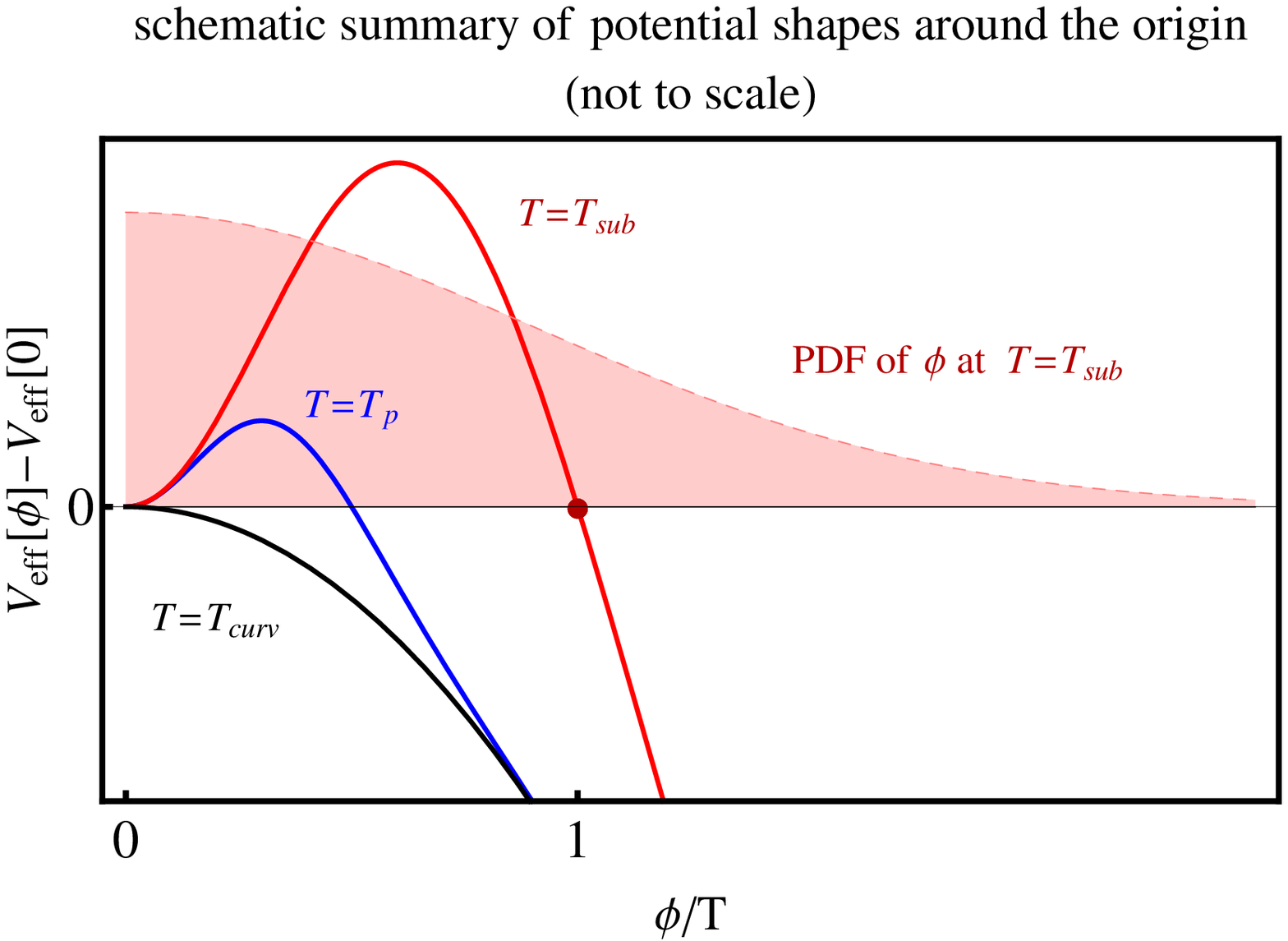}
 \end{center}
 \caption{A schematic relation of the potential shapes at $T=T_{\rm
 sub}$, $T_{\rm p}$, and $T_{\rm curv}$.  We also show the probability
 distribution function of the flaton at $T=T_{\rm sub}$, which indicates
 that subcritical bubbles are abundant in the Universe at $T=T_{\rm
 sub}$.  } \label{schematicshapeofV}
\end{figure}

By performing numerical simulations at $T=T_{\rm sub}$ we are able to
verify that the height of the potential barrier is small enough for the
flaton to escape from $\phi=0$.  In some cases we found that the
flaton rolls down to the bottom of the potential -- meaning that thermal
inflation ends at $T>T_{\rm sub}$ -- and in other cases we found that
the flaton remained around the origin, \footnote{This may be explained
as an effect of surface tension, which is stronger than the potential
force pulling the flaton away from the origin.} but with a distribution
width that was broader than the potential well.  We thus see that all
cases deviate from the standard scenario in which thermal inflation ends
as the result of a strong first-order phase transition.  We summarize
the dependence of the potential shape on temperature in
Fig.\ref{schematicshapeofV} schematically.

\begin{table}[h]
\scriptsize 
  \begin{tabular}{|c|c|c|c|c|c|c|} \hline \rule[0pt]{0pt}{8pt}
 \shortstack{ scenario \\{}} & 
 \shortstack{ couplings \\{}} &
 \shortstack{  $T_{\rm curv}$[GeV] \\{}}  & 
 \shortstack{  $T_{\rm p}$[GeV] \\{}} & 
 \shortstack{  $T_{\rm sub}$[GeV] \\{}} &
 \shortstack{  $F(T_{\rm sub})$ \\{}} &
  \shortstack{ simulated  $\sqrt{\langle \phi^2 \rangle}$ \\ at $T_{\rm sub}$ }\\ \hline \rule[0pt]{0pt}{10pt}
   MD $\to$ TI &strong & 5230 & 5239 ($2\times 10^{-3}$)& 5502($5\times 10^{-2}$) &  $10^{-84}$& $\phi_{\rm vev}$ \\\hline \rule[0pt]{0pt}{10pt}
   MD $\to$ TI &weak & 41216.96 &41216.97($4\times 10^{-7}$) & 41378($4\times 10^{-3}$) & less than $10^{-2000} $& $0.91T$\\\hline\rule[0pt]{0pt}{10pt}
   RD $\to$ TI &strong & 5230 & 5252 ($4\times 10^{-3}$)& 5502($5\times 10^{-2}$) &  $10^{-77}$& $\phi_{\rm vev}$  \\\hline\rule[0pt]{0pt}{10pt}
   RD $\to$ TI &weak & 41216.96 &41216.97($4\times 10^{-7}$) & 41378($4\times 10^{-3}$) & less than $10^{-2000} $ & $0.91T$\\\hline
  \end{tabular}
  \caption{Specific temperature values for four different scenarios.  In
  all four scenarios we take $m_{\phi}=1\,{\rm TeV}$.  Since the
  temperatures themselves are almost the same, we also show the relative
  differences $(T_{\rm p}-T_{\rm curv})/T_{\rm curv}$ and $(T_{\rm
  sub}-T_{\rm curv})/T_{\rm curv}$ in brackets.  In evaluating $T_{\rm
  p}$ and $F(T)$, we fix the value of $V_{\rm TI}$ so that thermal
  inflation begins at $T=T_{\rm curv}\times e^6$.  The RMS values of
  $\phi$ at $T=T_{\rm sub}$, obtained by simulations with duration
  $t=2000/T$, are also shown.  In the two strongly-coupled cases the
  flaton leaves the origin and settles in its VEV.  In the two
  weakly-coupled cases the flaton stays at the origin, but the width of
  its distribution function is as broad as the barrier.  Though the
  potential barrier is negligible, the potential force arising from the
  tachyonic mass term is also so weak that it may take a long time to
  displace the flaton from the origin.  } \label{temptable}
\end{table}

\section{Conclusion \label{conclusion}}

In this paper we studied the effect of thermal fluctuations on the
thermal inflation scenario.  Thermal inflation is a short period of
accelerated expansion after reheating and provides a way to dilute
dangerous moduli and gravitinos in order to make theories based on
supersymmetry compatible with cosmological observations.  Thermal
inflation is driven by the flaton potential energy at the origin with
the help of thermal corrections.  Since the thermal environment gives
rise to thermal fluctuations as well, we used lattice simulations to
study the dynamics of the flaton taking into account the 1-loop
effective potential, thermal fluctuations and the dissipation term.
First we studied the effects of thermal fluctuations before thermal
inflation.  Though the effective potential contains multiple local
minima during the course of the evolution of the Universe, the flaton
settles at the origin before thermal inflation even when thermal
fluctuations are taken into account.  Therefore the scenario of thermal
inflation may be feasible.  Second, we find that thermal inflation ends
with a cross over phase transition.  The tunneling rate of the
flaton from the origin of the potential is so small that the tunneling
does not occur until the position of the potential barrier becomes very
close to the origin.  However, since the height of the barrier is much
smaller than $T^4$, the flaton can escape over the barrier before
tunneling occurs.  Though the form of the effective potential suggests that
thermal inflation ends with a first-order phase transition accompanied
by bubble formation, thermal fluctuations make the transition 
to proceed through phase mixing,
which is characterized by subcritical bubbles.  
As such, we cannot expect critical bubble formation and the production of
gravitational waves.

\section*{Acknowledgments}
We would like to thank Jonathan White for helpful comments.  Y.M. also
thanks Kohei Kamada and Daisuke Yamauchi for informative discussions.
This work was supported in part by MEXT SPIRE and JICFuS (T.H.), JSPS
Research Fellowships for Young Scientists (Y.M.), and JSPS Grant-in-Aid
for Scientific Research No.23340058 (J.Y.)

\section*{Appendix : constructing approximation functions of the potential term}

In this section, we consider the approximation of Eq.~(\ref{eq:def_J}),
which determines the functional shape of the thermal correction to the
flaton potential. Expanding the integrand of Eq.~(\ref{eq:def_J}), we
can perform the integration term by term,
%
\begin{align}
J_\pm(y) &= \mp\frac{1}{2\pi^2}\sum_{n=1}^\infty\frac{(\pm 1)^n}{n}
   \int_0^\infty\!dx\, x^2 e^{-n\sqrt{x^2+y^2}}, \notag \\
&= \mp\frac{y^2}{2\pi^2}\sum_{n=1}^\infty\frac{(\pm 1)^n}{n^2}
   K_2(ny),
\label{eq:int_J}
\end{align}
%
where $K_2(x)$ is the modified Bessel function of the second kind. 
The derivative of $J(y)$ with respect to $y$,
which appears in the field equation, (\ref{simpleEOM}), is calculated as
%
\begin{align}
 \frac{dJ}{dy} &= 
   \pm \frac{y^2}{2\pi^2}\sum_{n=1}^\infty\frac{(\pm 1)^n}{n}K_1(ny).
\end{align}
%

For convenience, we define the shape function,
%
\begin{align}
 S_{\pm}(y) \equiv \sum_{n=1}^{\infty}\frac{(\pm 1)^n}{n}K_1(ny).
 \label{eq:shape}
\end{align}
%
The modified Bessel function $K_1(z)$ for small $z$ can be approximated as
%
\begin{align}
 K_1(z) \approx \frac{1}{z}.
 \label{eq:K1_small}
\end{align}
%
Therefore, the shape function for small $y$ becomes
%
\begin{align}
 S_{\pm}(y) \approx 
   \frac{1}{y}\sum_{n=1}^{\infty}
   \frac{(\pm 1)^n}{n^2}
 =\frac{1}{y}\times
\begin{cases}
\displaystyle \zeta(2), & \mbox{for }+\\
\displaystyle -\frac{\zeta(2)}{2}. & \mbox{for }-
\end{cases}
\label{eq:small_z}
\end{align}
%
Away from $y=0$ this approximation breaks down almost
immediately. Moreover, it is difficult to achieve better accuracy by
simply retaining more terms in the expansion in Eq.~(\ref{eq:K1_small}),
since there are logarithmic terms like $\ln z$, meaning that we cannot
take the infinite summation analytically. Instead, we use the following
ansatz,
%
\begin{align}
 \widetilde{S}^{(0)}_{+}(y) &=
 \frac{e^{-y}}{y}\left(\zeta(2) + a_1y + a_2y^2 + a_3y^3\right),\\
 \widetilde{S}^{(0)}_{-}(y) &=
 \frac{e^{-y}}{y}\left(-\frac{\zeta(2)}{2} + b_1y + b_2y^2 + b_3y^3 + b_4y^4\right),
\end{align}
%
where $a_i$ and $b_i$ are determined by requiring a good fit with the
shape function in the limited region $0\leq y\leq 2$; we obtain
$a_i=(0.146773, 0.106023, -0.0248936)$ and $b_i=(-0.772073, 0.163142,
-0.0547415, 0.0107667)$.

In the opposite limit, for large $y$ we can truncate the infinite summation in Eq.~(\ref{eq:shape}) 
at relatively small $n$ thanks to
the asymptotically exponential decay of $K_1(ny)$. Here we take the
summation up to $n=2$. We also use  
the asymptotic expansion of the modified Bessel
functions. To guarantee accuracy, we expand $K_1(y)$ up to $y^{-3}$ and
$K_1(2y)$ up to $y^{-1}$. Eventually we obtain
%
\begin{align}
 \widetilde{S}^{(\infty)}_{\pm}(y) =
 \pm\sqrt{\frac{\pi}{2y}}e^{-y}
    \left(1+\frac{3}{8y}-\frac{15}{128y^2}+\frac{105}{1024y^3}\right)
 +\sqrt{\frac{\pi}{16y}}e^{-2y}\left(1+\frac{3}{16y}\right).
\end{align}
%
Finally, we approximate the shape function given in Eq.~(\ref{eq:shape}) as
%
\begin{align}
 S_\pm(y) \approx
\begin{cases}
 \widetilde{S}_\pm^{(0)}(y), & {\rm for}\;y<2, \\[2mm]
 \widetilde{S}_\pm^{(\infty)}(y), & {\rm for}\;y \geq 2.
\end{cases}
\end{align}
%
The partitioned fitting curve for the shape function constructed here
has an accuracy $E=1.73\times 10^{-3}$ for $S_{-}$ and $E=2.06\times
10^{-3}$ for $S_+$, where $E\equiv
||1-\widetilde{S}_{\pm}(y)/S_{\pm}(y)||_\infty$.  Note that, as a result
of the naive matching of the two functions, $dV^{1{\rm -loop}}_T/d\phi$ is
discontinuous at $y=2$ by construction.  However, this is not
problematic, since the amplitude of the discontinuity in $dV_T/d\phi$ at
$y=2$ is on the order of $0.1\%$.

\end{document}